# IMAGE SUBTRACTION NOISE REDUCTION USING POINT SPREAD FUNCTION CROSS-CORRELATION

*Steven Hartung*

s.hartung@computer.org

**ABSTRACT**

Image subtraction in astronomy is a tool for transient object discovery and characterization, particularly useful in wide fields, and is well suited for moving or photometrically varying objects such as asteroids, extra-solar planets and supernovae. A convolution technique is used to match point spread functions (PSFs) between images of the same field taken at different times prior to pixel-by-pixel subtraction. Particularly suitable for large-scale images is a spatially-varying kernel, where the convolution is allowed to adapt to PSF changes as a function of position within the images. The most versatile basis for fitting the spatially-varying kernel is the Dirac delta function. However, the convolution kernel based on the delta function does not discriminate between pixel scale noise variations and the intended stellar point spread function signals. The situation can frequently lead to reduced signal to noise ratios for variable objects detectable in the resulting subtraction. This work presents a cross-correlation method for reducing noise effects on the delta function derived convolution kernels, thus yielding significantly improved signal to noise in the resulting subtraction.

## 1. INTRODUCTION

The optimal image subtraction (OIS) method matches point spread functions (PSFs) between two images via a convolution of one image to match the other, where the convolution kernel is determined by a fitting process (Alard, 2000; Alard et al., 1998). The original OIS implementations relied on modeling the convolution kernel as a superposition of Gaussian function basis (GFB) components. An alternate method that is more adaptable to arbitrary PSF forms is a superposition of Dirac delta function basis (DFB) components (Miller et al., 2008). Unfortunately, it has been shown that the DFB in a spatially-varying kernel OIS will tend to "over-fit" the kernel, responding to image noise components as much as the desired signal (Becker et al., 2012). The side-effect of kernel over-fitting is to increase noise in the final subtraction result, reducing the detection capability for small photometric changes. In order to reduce over-fitting, Becker et al. have applied constraints to the DFB kernel based on statistical methods. This work presents an alternate approach which dramatically reduces the noise components prior to DFB kernel fitting.

Cross-correlation is a standard signal processing method often used for extracting a signal of known form from a noisy environment (Oppenheim et al., 1983). The use of correlation is extremely powerful for improving SNR prior to convolution kernel fitting. Most variable object studies are searching for, or monitoring, objects that appear as unresolved point sources. This fact allows the use of the stellar PSF as a model for the expected signal and provides a correlation basis. The cross-correlation method uses an expected signal, in this case the modeled PSF, as opposed to the auto-correlation where an entire image is correlated to itself. The cross-correlation of the expected signal against an image is shown in equation(1).

$$I_{x,y} \star P_{x,y} = \sum_{u=x-w}^{x+w} \sum_{v=y-w}^{y+w} I_{u,v} P_{u-x+w+1, v-y+w+1} \qquad (1)$$

Where $\star$ represents the correlation operator, *I* is the image pixels, and *P* represents the PSF model pixels.

## 2. CONSTANT PSF MODEL

OIS determines PSF changes from qualified sample stars distributed across the image. The samples regions have been given the term stamps, each stamp consisting of a star and a small region of sky around it. Ideally, the stamps contain isolated single stars. All samples must be unsaturated and within the linear response range of the detector. In OIS the stamps are evaluated to determine the necessary convolution to match the stars in the two images. The selection and qualification of the stamps also provides an excellent data set for identifying a sample PSF in each individual image. From these stamps it is possible to create a PSF model from the image data itself.

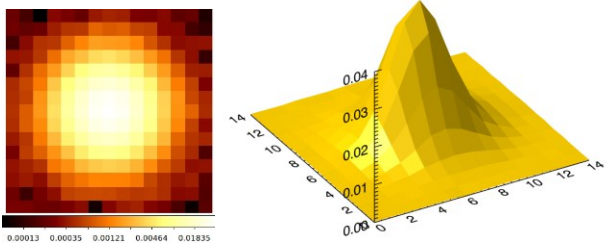

Figure 1 Example PSF model derived from an image in 2-D and 3-D representation.

For the cross-correlation, it is necessary to identify expected signal form. This can be accomplished by first normalizing all stamps to a specified constant integrated flux. First, any background sky offset must be subtracted, after which the PSF amplitude can be scaled to a standardized value. For convenience sake, a scaling to unity net flux is chosen here, though the actual value is somewhat arbitrary. Once all stamps have been scaled to the same flux, an expected signal for the entire image can now be found as a simple average of all stamps. Once the expected PSF has been determined, a correlation operation will enhance all image elements matching the expected signal, and will smooth all features that do not match the expectation.

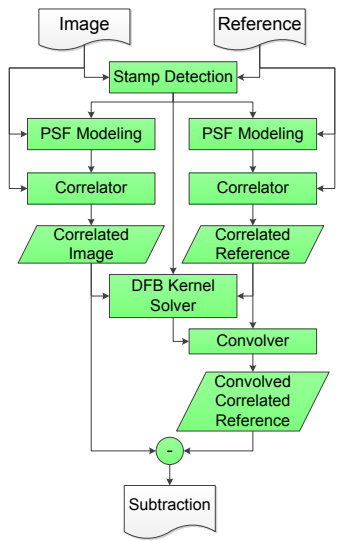

Figure 2 Process flow for OIS modified with PSF correlation.

For a spatially-varying DFB OIS it is useful to perform the correlation matching prior to the kernel fitting in order to reduce pixel scale noise that the kernel may otherwise inherit. The resulting subtraction demonstrates a significant improvement in SNR, thus improving the target object's detectability by automated means.

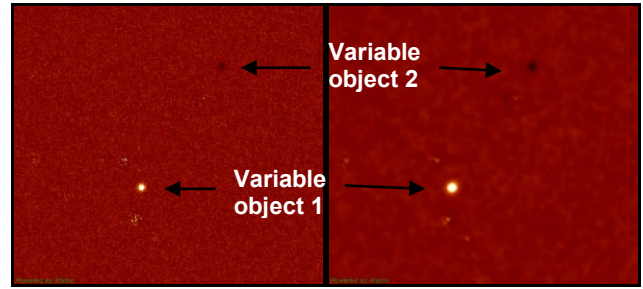

Figure 3 Example of standard OIS subtraction on left, and subtraction utilizing PSF cross-correlation preprocessing on the right. The same two variable objects can be seen in both examples, however the image on the right demonstrates improved SNR.

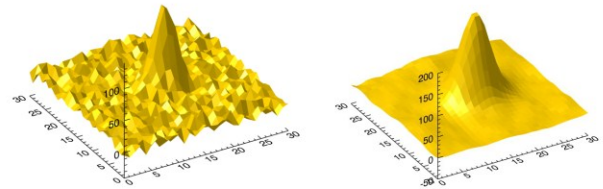

Figure 4 A close-up view of variable object 1 from Figure 3 shown in 3-D relief illustrating the improved SNR. The image on the right demonstrates the PSF cross-correlation preprocessed subtraction.

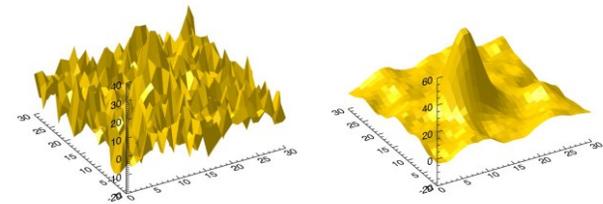

Figure 5 A close-up view of variable object 2 from Figure 3 shown in 3-D relief illustrating the improved SNR. The image on the right demonstrates the PSF cross-correlation preprocessed subtraction.

The examples shown in Figures 3-5 demonstrate an improvement in SNR in the subtraction result of approximately 2.6x. Variable object 2 is of particular interest, in the original subtraction it is rather challenging to pick out the object from the noise. In contrast, the PSF cross-correlation preprocessed subtraction has dramatically differentiated the object from the background.

## 3. SPATIALLY-VARYING PSF MODEL

In some cases, particularly in a very wide FOV, OIS has demonstrated that the PSF can vary significantly across the image. Thus, the cross-correlation should also be adaptable to PSF changes in order to achieve maximum detection. The spatially-varying cross-correlation uses the same least-squares fitting philosophy as spatially-varying OIS. As before, the stamps must be normalized to a constant flux. Instead of averaging all of the stamps in the image, it is now useful to produce running average models for each region of the image. In this way the model avoids the original pitfall of the OIS DFB over-fitting the kernel. A least-square fit of the PSF changes then produces an expected signal that evolves as a function of position in the images.

As in the case of OIS, the evolution of the PSF is assumed to be due to changes in focus and optical path, both in the atmosphere and in the instrument, and is thus considered to be a smooth and continuous change.

In the case of the spatially-varying PSF cross-correlation,

$$F_{u,v} = \sum_{k=1}^{S} \left( \left( a_n(x,y) \right)_{u,v} - \bar{P}_{k,u,v} \right)^2 \quad (2)$$

Where $u,v$ indicates the pixel position in the regional running average PSF represented by $\bar{P}$, and $S$ is the number of stamps. A bivariate polynomial of order $n$ is created for each pixel in the PSF, and the coefficients of $a_n$ describe the value of the pixel $u,v$ in PSF over the entire image. For the second order polynomial, $a_n$ takes the form of equation (3).

$$a_n(x,y) = a_{00} + a_{01}y + a_{02}y^2 + a_{11}xy + a_{10}x + a_{20}x^2 \quad (3)$$

The spatially-varying PSF is obtained by solving for the coefficients of $a_n$ for each PSF pixel such that $\nabla F$ is minimized.

## 6. CONCLUSIONS

The DFB over-fitting in spatially-varying OIS is related to the noise in the images effecting the kernel determination. One approach to mitigating the effects of over-fitting is to reduce the pixel-scale image noise prior to the kernel fitting process. PSF cross-correlation offers a method for improving SNR for unresolved point source objects. The noise reductions that are achieved carry through to the kernel fitting process and the final subtraction.

In many cases a constant PSF model may be sufficient for an entire image. In the cases where a single PSF model is not sufficient, a spatially-varying PSF model can be created by similar means to the fitting process used for the spatially-varying convolution kernel in OIS.

The method is not well suited to extended and resolved sources such as planetary imaging or variable nebula since it relies on a PSF model as generated by a point source sample for the expected signal. However, most active areas of transient study in astronomy are addressing point source objects, for which the method is very well suited.

While the cross-correlation does cause a blurring of the image, it may still be desirable to apply this method to processing beyond OIS, where improving the signal to noise is of greater importance than sharp focus.